# The high-performance data acquisition system for the GAMMA-400 satellite-borne gamma-ray telescope

**A.V. Bakaldin[1][ab], S.G. Bobkov[a], O.V. Serdin[a], M.S. Gorbunov[a], A.I. Arkhangelskiy[c], A.A. Leonov[cb], and N.P. Topchiev[b]**

[a] *Scientific Research Institute for System Analysis of the Russian Academy of Sciences (SRISA)*
  *Nakhimovskiy prospect 36/1, Moscow, 117218, Russia*

[b] *Lebedev Physical institute of the Russian Academy of Sciences*
  *Leninskiy prospect 53, Moscow, 119991, Russia*

[c] *National Research Nuclear University MEPhI (Moscow Engineering Physics Institute)*
  *Kashirskoe highway 31, Moscow, 115409, Russia*
*E-mail:* `bakaldin71@mail.ru`

The future GAMMA-400 space mission is aimed for the study of gamma rays in the energy range from ~20 MeV up to 1 TeV. The observations will carry out with GAMMA-400 gamma-ray telescope installed on-board the Russian Space Observatory. We present the detailed description of the architecture and performances of scientific data acquisition system (SDAQ) developing by SRISA for the GAMMA-400 instrument. SDAQ provides the collection of the data from telescope detector subsystems (up to 100 GB per day), the preliminary processing of scientific information and its accumulation in mass memory, transferring the information from mass memory to the satellite high-speed radio line for its transmission to the ground station, the control and monitoring of the telescope subsystems. SDAQ includes special space qualified chipset designed by SRISA and has scalable modular net structure based on fast and high-reliable serial interfaces.

---

[1]Speaker

## 1. Introduction

The future GAMMA-400 space mission [1] is aimed for the study of gamma rays in the energy range from ~20 MeV up to 1 TeV. The observations will carry out with GAMMA-400 gamma-ray telescope installed on-board the Russian Space Observatory.

The physics scheme of the GAMMA-400 gamma-ray telescope is presented in figure 1. The GAMMA-400 instrument is composed by the following main subsystems: anticoincidence subsystem (AC), converter-tracker subsystem (C), a Time of flight subsystem (TOF), an imaging calorimeter (CC1), deep electromagnetic calorimeter (CC2), trigger system (Tr), scientific data acquisition system (SDAQ), and telescope power supply subsystem (PSS). The detailed description of the GAMMA-400 telescope can be found in [2]. The volume of data acquired from the instrument subsystems for separate detected event is not exceeded 1 MByte. The estimated maximum volume of scientific information collected by the telescope per day is about 100 Gbytes.

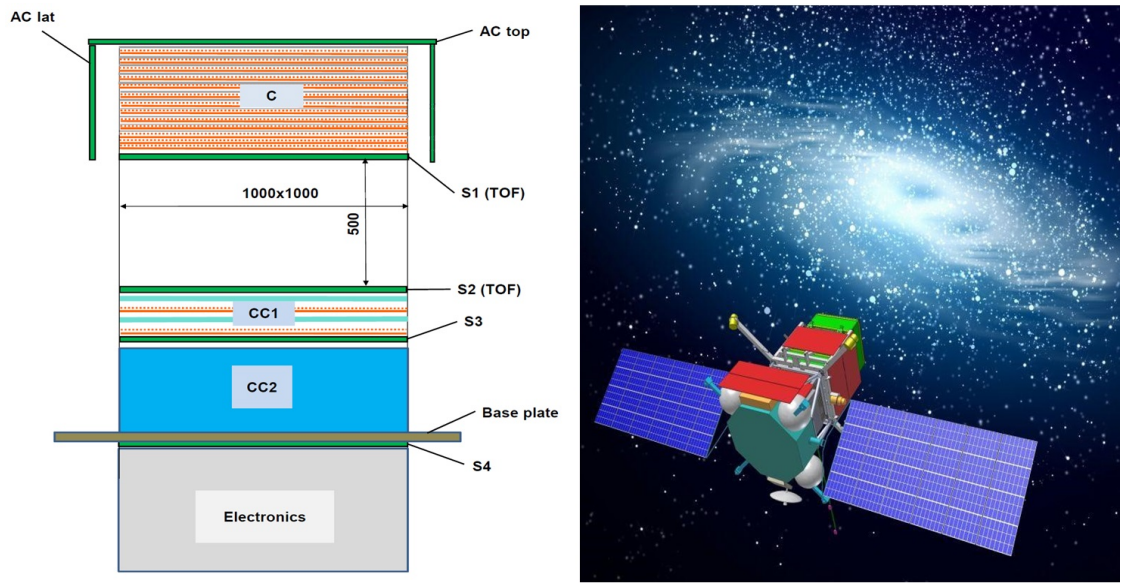

**Figure 1:** The GAMMA-400 telescope physics scheme (left) and artistic view of the GAMMA-400 Space Observatory (right). The instrument trigger system (Tr), scientific data acquisition system (SDAQ) and power supply subsystem (PSS) are installed inside the Electronics unit.

The SDAQ is the heart of scientific complex and so it should be a high-reliable subsystem. It provides the instrument control, the acquisition, pre-processing and accumulation in mass memory of the scientific and housekeeping data from telescope detector subsystems, transferring the collected data to the satellite radio line. The SDAQ architecture proposed in this work allows us to realize the effective control and fast scientific data acquisition for the GAMMA-400 instrument. In order to increase the reliability of SDAQ, it is designed using a scheme with reserved subsystems. All control and data transfer interfaces are double redundant. Additional reliability level of SDAQ is achieved by minimization of the number of high integrity chips.

## 2. The main SDAQ functions

The main functions of the scientific data acquisition system are the following:
- The data acquisition from subsystems of the gamma-ray telescope.
- Preliminary processing of scientific information and storage it in non-volatile mass memory (1 TByte total).
- Scientific data transfer into high-speed (320 Mbit/s) scientific radio line (SRL) for its transmission to the data acquisition ground stations.
- Control information reception from the satellite on-board control system (OCS) via MIL-STD-1553B interface, its decoding and transfer to telescope subsystems. Acquisition of housekeeping data and their transmission to OCS.
- Receiving signals from on-board time (OBT) and frequency standard system and on-board control system and generating high-stable reference synchronization signals and instrument time code for precise timing of telescope subsystems.

## 3. The description of the GAMMA-400 scientific data acquisition system

The functional diagram of SDAQ is shown in figure 2. The SDAQ consists of power supply control module (BTSSNI-001) and two identical subsystems: SDAQ Sub main and SDAQ Sub spare.

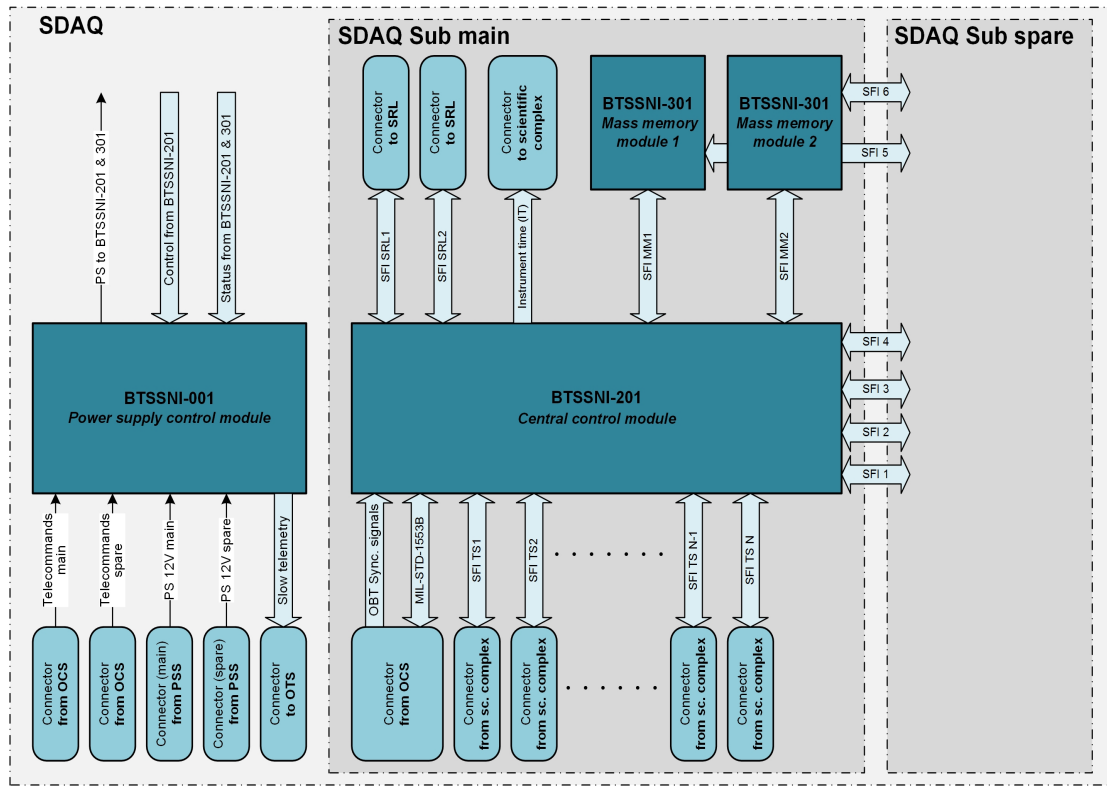

**Figure 2:** Functional diagram of GAMMA-400 scientific data acquisition system (OCS is on-board control system, PSS is telescope power supply system, OTS is on-board telemetry system, SRL is scientific radio line, SFI is serial fast interface).

Each redundant subsystem includes two types of modules: central control module (BTSSNI-201) and mass memory module (BTSSNI-301).

The main functions of BTSSNI-001 are the filtering of the main and spare power supply channels (12 V) received from the telescope power supply system and the power supply switching-on/off for the main/spare BTSSNI-201 and BTSSNI-301 modules.

The central control module BTSSNI-201 is the intelligent core of SDAQ. It executes the receiving of macrocommands from OCS and distributing of the control information to telescope subsystems, the scientific and housekeeping data acquisition from the subsystems and their distribution between two mass memory modules BTSSNI-301 n.1 and BTSSNI-301 n.2. BTSSNI-201 generates high-stable reference synchronisation signals (1 Hz and 1 MHz) and 32-bit instrument time code for precise timing of scientific complex. For the timing with UT BTSSNI-201 module periodically receives the synchronisation signals and onboard time code from the on-board time and frequency standard system and from the on-board control system. BTSSNI-201 controls the whole SDAQ functioning.

The BTSSNI-301 unit is intended for acquired data preliminary processing and storage in non-volatile mass memory. BTSSNI-301 contains 256 GBytes NAND flash memory bank. Two BTSSNI-301 modules are utilised in SDAQ to increase the speed of scientific data acquisition, the volume of collected information and reliability of the system.

All SDAQ modules with backplane are installed in special aluminium alloy case.

SDAQ has the following redundant external interfaces: pulse telecommand interface (from on board control system (OCS), power channel +12 V (from telescope power supply system (PSS)), slow telemetry channel (to on board telemetry system (OTS)), intelligent control interface MIL-STD-1553B (to OCS), up to N=16 serial fast interface SFI channels (the throughput of each channel is not less then 300 Mbit/s) for the control and scientific data acquisition from telescope subsystems, two SFI channels for the data transmission from mass memory into satellite scientific radio line SRL, time synchronisation interface (to the instrument subsystems).

The SDAQ technical characteristics are: the total maximum throughput of all SFI channels for scientific data acquisition from telescope subsystems is 70 MBytes/s; the maximum data rate of two SFI channels for the data transmission to SRL is 40 MBytes/s; mass memory volume is 1024 GByte; maximum power consumption is 80 W; outline dimensions are 400×250×240 mm; mass is not more then 24 kg.

The detailed design of BTSSNI-001, BTSSNI-201, BTSSNI-301 units, as well as the architecture of internal SFI network and description of the used software are presented below.

**3.1 The power supply control module BTSSNI-001**

The flowchart of power supply control module BTSSNI-001 is shown in figure 3. The core of BTSSNI-001 is redundant control logic unit and relay unit. BTSSNI-001 provides separate switching-on/off of BTSSNI-201, BTSSNI-301 main and spare modules. The switching-on/off of the BTSSNI-201 units is carried out by on-board control and telemetry systems with pulse telecommands, and the switching-on/off of the BTSSNI-301 modules is controlled by BTSSNI-201. Simultaneously the BTSSNI-001 can switch-on only the one of the main and spare module.

The proposed BTSSNI-001 organisation provides the full backup of SDAQ subsystems.

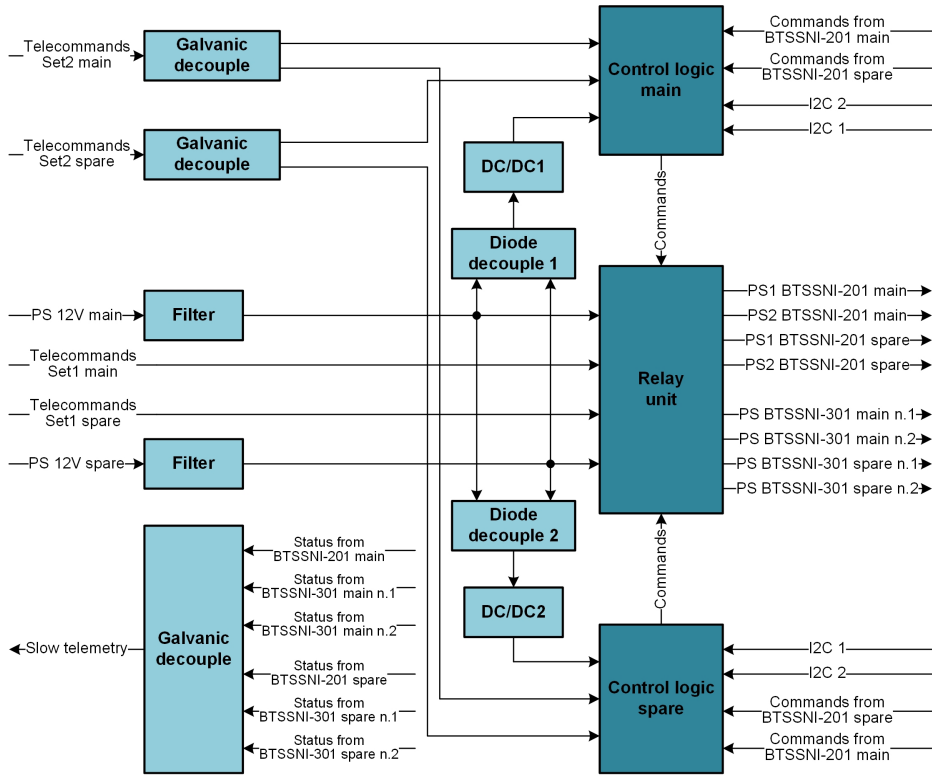

**Figure 3:** BTSSNI-001 scheme.

## 3.2 The central control module BTSSNI-201

The scheme of central control module BTSSNI-201 is shown in figure 4.

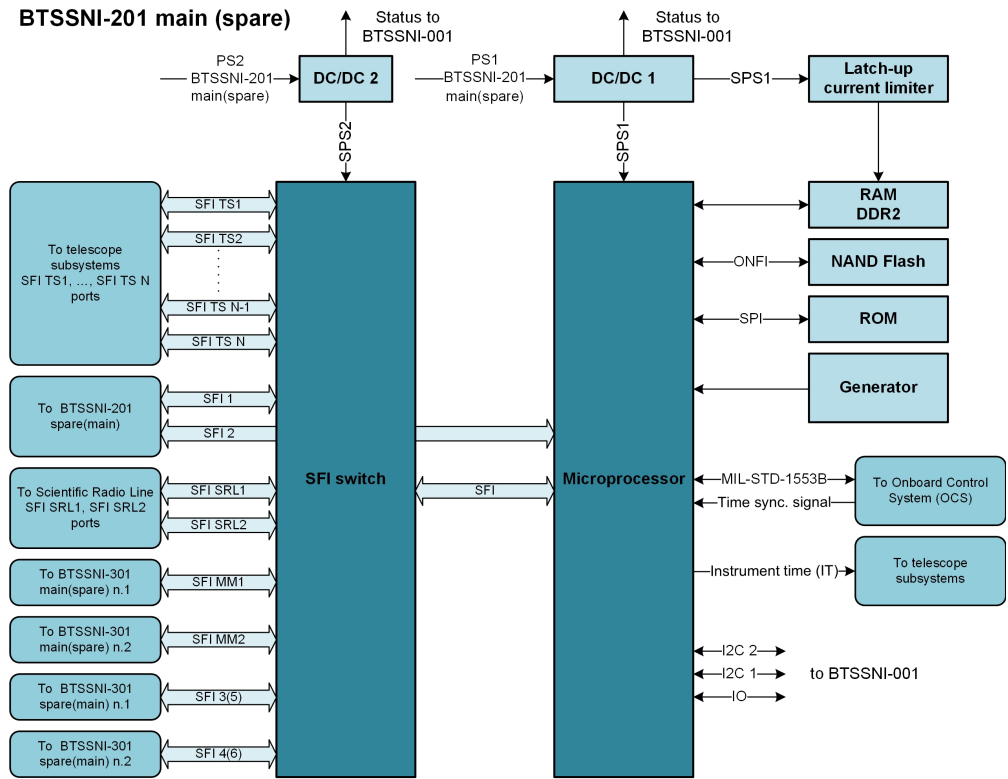

**Figure 4:** BTSSNI-201 scheme.

BTSSNI-201 contains system-on-chip microprocessor and serial fast interface SFI switch. System-on-chip microprocessor is based on processor core KOMDIV-32 [3-6]. It consists of 32-bit central processing unit (CPU), 128-bit arithmetic co-processor CP2, system controller with DDRII, SPI, SFI ports, I2C, GPIO, UART. Clock frequency is about 100 MHz, the throughput of RAM is 512 Mbytes/s. The SFI switch consists of SFI ports; the transferring environment is configured independently. Transferring speed is not less then 300 Mbit/s per channel.

### 3.3 The mass memory module BTSSNI-301

The BTSSNI-301 module contains system-on-chip microprocessor and 256 GByte NAND flash mass memory. The mass memory is divided in eight 32 Gbyte banks. The BTSSNI-301 detailed flowchart is presented in figure 5.

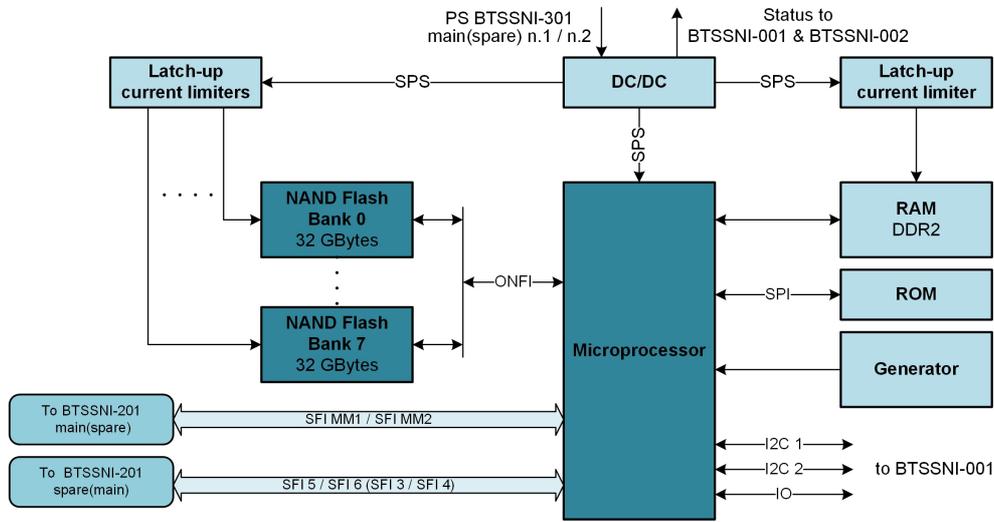

**Figure 5:** BTSSNI-301 scheme.

### 3.4 The architecture of internal SFI network

The architecture of SDAQ internal SFI network is shown in figure 6.

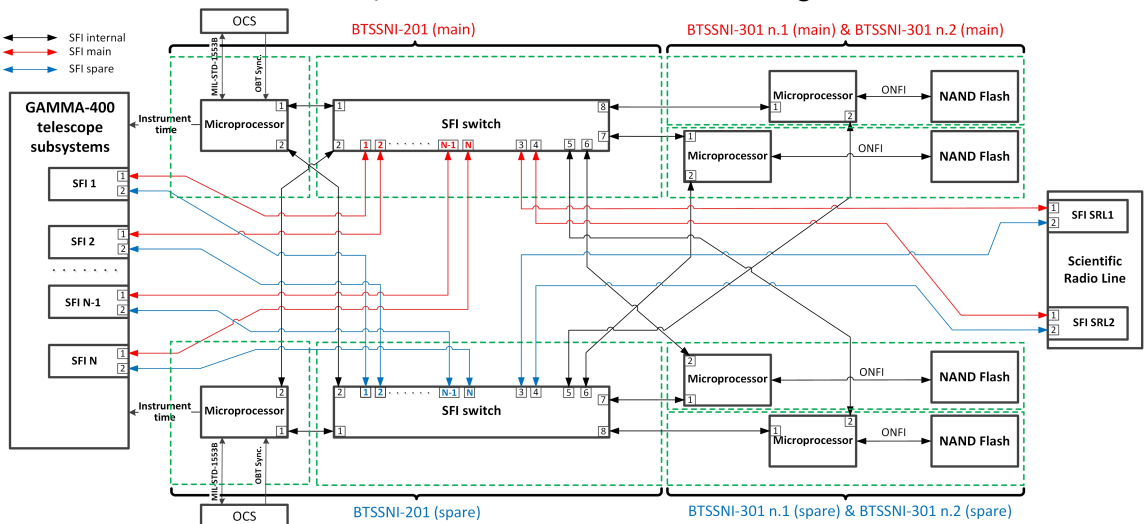

**Figure 6:** The architecture of SDAQ internal SFI network.

The control of the telescope and the data acquisition from its subsystems are provided with up to N=16 SFI channels (SFI 1-SFI N). Another two SFI interfaces (SFI SRL1, SFI SRL2) is used for the transferring of the data stored in NAND flesh mass memory to scientific radio line of the satellite. The throughput of each SFI channel is not less then 300 Mbit/s. The architecture of SFI network allows to achieve the full backup of SDAQ subsystems.

**3.5 The SDAQ real-time operational system**

As operating system for SDAQ, the real-time OS (RTOS) Baget 3.0 [7, 8] was chosen. This RTOS is developed on the base of the following general approaches:
- use of standards (ARINC 653 and POSIX 1003.1 for programming interface; C standard for C language and libraries);
- portability;
- advanced facilities for tracing, logging, diagnostics. and error handling (health monitor);
- flexible scheduling;
- object-oriented approach;
- scalability (configuration tools);
- instrumental software for developing and debugging user cross-applications;
- large number of environmental packages for creating graphics applications, databases, and mapping systems.

For the best portability, OS is divided on three main parts: main part independent on hardware, written on C language (the biggest); the second part, dependent only on central processor type, written on C or Assembler language (much smaller); modules support part containing modules drivers.

During software design, cross-development technology is used whereby, using the host computer with a general-purpose OS, the source codes and OS libraries are stored as well as the compilation and build of the boot image that is executed on the target computer under the control of RTOS are performed. For debugging in terms of the source code, a remote debugger is used.

Additional reliability level of SDAQ is achieved by minimization of the number of high integrity chips.

**4. Conclusion**

For developing of the GAMMA-400 SDAQ the modern technical solutions were utilised. Thereby SDAQ has cross-redundant and high reliable structure. The applying of parallel architecture in combination with serial fast interfaces for scientific information acquisition will allow us to decrease the instrument dead time and obtain up to 100 GBytes of experimental data per day.

At present stage of the GAMMA-400 project development the full-scale prototype of SDAQ is being designed for justification of the principle engineering solutions.

**References**

[1] N.P. Topchiev et al., *Journal of Phys.: Conference Series* **675** (2016) 032010.

[2] N.P. Topchiev et al., *Journal of Phys.: Conference Series* **675** (2016) 032009.

[3] S.G. Bobkov, *Information Technologies* **12 (**2012) 2 (In Russian).

[4] I.A. Sokolov et al., *Informatics and its Applications* **8(1)** (2014 ) 45 (In Russian).

[5] S.G. Bobkov, *Vestnik Rossiiskoi Akademii Nauk* **84(11)** (2014) 1010 (In Russian).

[6] S.G. Bobkov et al., *Software & Systems* **4** (2013) 49 (In Russian).

[7] A.N. Godunov, *Software & Systems* **4** (2010) 16 (In Russian).

[8] A.N. Godunov and V.A. Soldatov, *Programming and Computer Software* **40(5)** (2014) 259.